# A novel DL approach to PE malware detection: exploring Glove vectorization, MCC_RCNN and feature fusion

Yuzhou Lin[1]

[1] Beijing Key Laboratory of Security and Privacy in Intelligent Transportation, Beijing Jiaotong University, China

**Abstract.** In recent years, malware becomes more threatening. Concerning the increasing malware variants, there comes Machine Learning (ML)-based and Deep Learning (DL)-based approaches for heuristic detection. Nevertheless, the prediction accuracy of both needs to be improved. In response to the above issues in the PE malware domain, we propose the DL-based approaches for detection and use static-based features fed up into models. The contributions are as follows: we recapitulate existing malware detection methods. That is, we propose a vectorized representation model of the malware instruction layer and semantic layer based on Glove. We implement a neural network model called MCC_RCNN (Malware Detection and Recurrent Convolutional Neural Network), comprising of the combination with CNN and RNN. Moreover, we provide a description of feature fusion in static behavior levels. With the numerical results generated from several comparative experiments towards evaluating the Glove-based vectorization, MCC_RCNN-based classification methodology and feature fusion stages, our proposed classification methods can obtain a higher prediction accuracy than the other baseline methods.

## 1 Introduction

In recent years, the computer network has become a crucial foundation for the development, and network security issues have received unprecedented attention [1]. In the increasingly saturated Internet era, the number of malware is also increasing, and their attacking methods are also becoming polymorphic constantly [2]. In the "Global Risk Report 2018" released by the World Economic Forum, cyberattacks were declared for the first time as the third largest risk factor in the world [3].

As mentioned above, the number of malware is increasing recently, and many malware families underwent a structurally perfect evolution. The degree of malware damage is increasing, which becomes a national security strategy that cannot be ignored. In order to handle with the hardy problem of unknown malware variants, a trend of using ML-based models have improved but still lack of total raw data processing. Among all the models, DL models can convert detection issues into some problems in other domains, and get high prediction accuracy. However, most work focused on DL-based detection haven't been fully explored about their structure and optimization. Hence, optimizing model architecture and selecting the effective features is crucial in the future work.

Compared with the previous work focused on DL-based detection, diversified features have been used and different DL models have been embedded. In regard to the challenge of optimizing the current detection architecture of DL models, this paper seeks to give a novel approach for better malware detection accuracy based on DL models. We overview the related work on DL-based malware detection and then focus on the feature vector representation method of malware, feature extraction, feature fusion, and classification model towards detection. The overall architecture proposed in our approach is illustrated in **Fig**. 1:

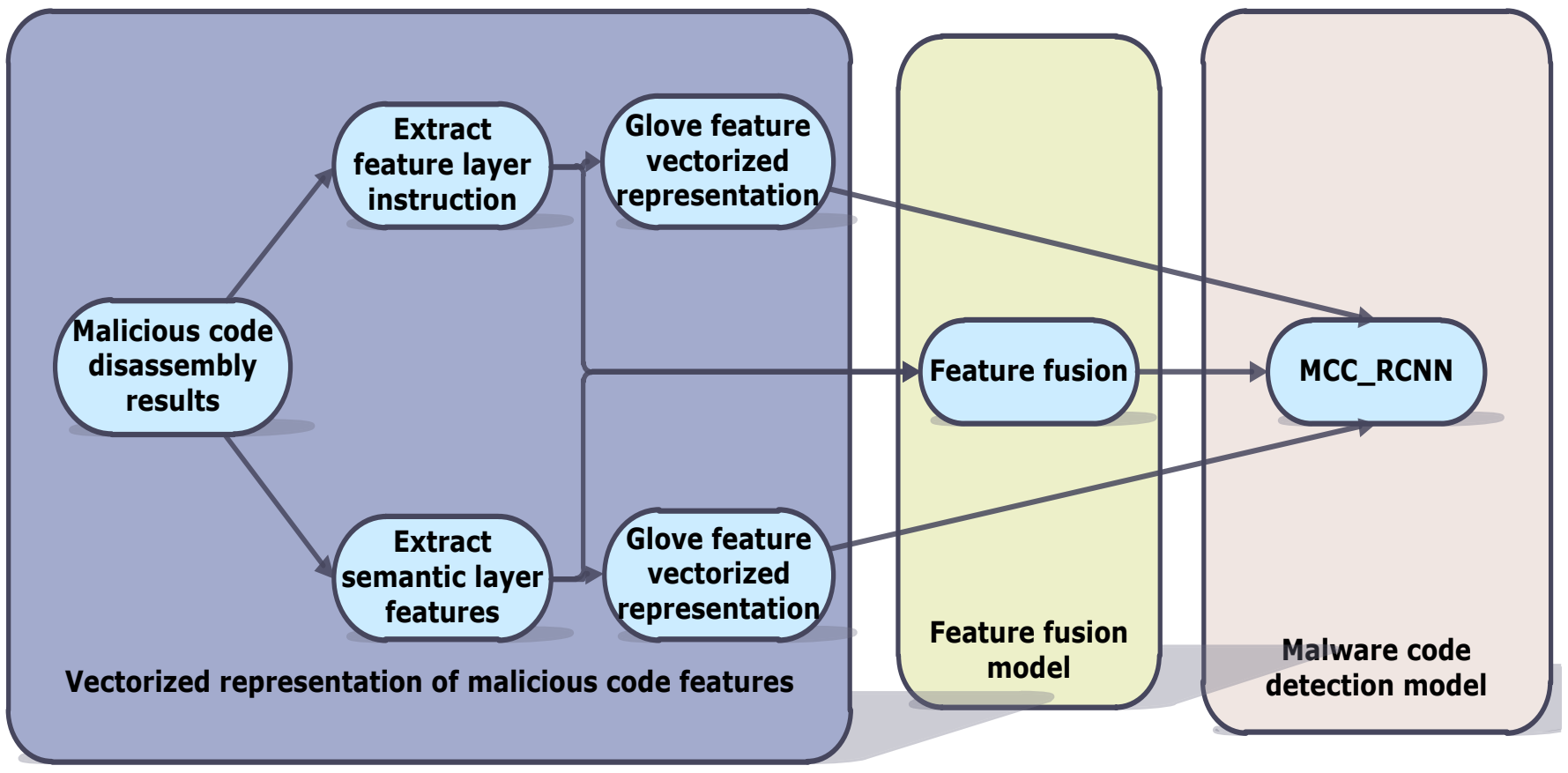

**Fig**. 1 The architecture of our approach

The contributions of this paper are summarized as follows:

(1) We propose a vectorized representation model of the malware instruction layer and semantic layer based on Glove. Concretely, we start with the static analysis to extract opcode and APIs these semantic features and evaluate the Glove-based feature processing comparing with n-gram window sliding approaches.

(2) A feature description method for feature fusion at the static behavior level is proposed. The results of the detection are largely dependent on the characterization method of malware. The characteristics of the instruction layer and the semantic layer can well represent the behavior information. To take full advantage of static analysis and reduce the impact of static analysis code confusion, we combine the instruction layer and semantic layer features of malware.

(3) We propose a malware detection model MCC_RCNN (Malware Detection_ Recurrent Convolutional Neural Network) on the basis of CNN and RNN. MCC_RCNN integrates LSTM and Gated CNN, which can not only solve the problem that the LSTM model cannot extract sequence feature information on a too long length scale, but also eliminate the problem of opcode association where the features of the CNN model alone do not have context.

The remainder of the paper is organized as follows. Section 2 overviews the related work about the confidential model for sequence processing called LSTM as well as the crucial semantic malicious features called API and recapitulates the related work towards DL-based malware detection. In section 3, we introduce the main principles of our methodology, which consists of Glove vectorization, MCC_RCNN, and feature fusion. In section 4, we implement several comparative experiments to evaluate the effectiveness of the parts, as mentioned earlier in this approach. Section 5 briefly provides the conclusion and prospect of future work.

## 2 Related work

### 2.1 Feature extraction

Feature extraction stage involves with selecting the effective features from malware and develop architectures to extract them. In general, there are instruction layer feature and semantic layer feature for extraction.

The instruction layer feature extraction of malware is to extract the operation code (Opcode) sequence of the instruction layer by the statistical model language algorithm on the disassembly result of the malware. Each instruction uses an opcode to tell the CPU program what kind of operation to perform. Disassembly translates opcodes into human-readable instructions. PE executable files can be disassembled to obtain ASM files, which are composed of assembly code. Then a series of opcodes at the instruction level can be extracted using some algorithms that express features. This technique was first proposed in [19]. The complete machine language instructions include operation codes and operands, and the operation codes operate according to the operands. The operation code is fixed in the computer. Due to the different nature of the operation, it includes logical operations and arithmetic operations. In PE malware, common instruction operation codes are push, XOR, mov, call, add, sub, etc.; the operand is the operation frequency of the operation code. The operands vary greatly depending on the operation target. The opcode sequence represents the execution logic and flow of malware, so the opcode has the characteristics of timing relationship.

Bilar et.al [20] used frequency statistical analysis on the distribution of opcodes to detect malware. It is found that the distribution of malware opcodes is statistically different from that of non-malware software. It proves that the rare opcode seems to be a more powerful predictor, which can explain the frequency change of 12% to 63%. They confirmed that it can extract features through opcode instructions and operands to detect malware. They also pointed out that the opcode and operand from the malware code will be different from the normal code. The malware will call some uncommon opcode. Therefore, the opcode is a feature that can better represent malicious behavior information in static analysis. Through the operation code, you can know what operations are in the assembly code of the executable file, and the PE malware is detected by the order and frequency of these operations.

Extracting the semantic layer features is to selecting and extracting the API calls generated from malware samples. API is called application program interface, used by Windows kernel to operate functions and command the operating system to work on specific tasks on the layer of user or kernel. In another word, API calls traces can reflect the malicious behaviors instead of the original code from files. API is divided into static API and dynamic API. The static API is obtained by static analysis in the format of ITA. On the one hand, we can use PE analysis tools to obtain the DLL call sequence in the import table. On the other hand, we can disassemble the malware and extract the API call sequence by analyzing the disassembly code. The dynamic API is a run-time API call sequence obtained by actually running malware samples in a sandbox such as Cuckoo. The basic behavior operations include process operations, file operations, registry operations, system calls, network operations, and memory management, etc.

The current research on the feature extraction of malware mainly has the following problems: 1. the dynamic analysis of malware and the extraction of features are subject to the constraints of the execution environment, the malware program execution path is single, and the behavioral characteristics of the full path of the malware cannot be obtained, and the overhead of dynamic analysis is large and the efficiency is low. 2. the traditional feature extraction method only extracts the features of the malware in the text layer of the malware, lacking the description and similarity of the contextual behavior information, and may be subject to code confusion. 3. the traditional ML classification feature extraction methods usually stay at the surface, which refers to that the extracted features are often shallow features. These models cannot effectively extract the deeper features, resulting in a lower accuracy of detection results.

## 2.2 Confidential DL models

After feature extraction, we should use some models to process the sequences. LSTM is a commonly used RNN that can learn long-term dependencies. It was proposed by Hochreiter et al. [21] and later refined and promoted by many researchers in their work. LSTM stands out on various issues and is currently widely used in malware detection. The purpose of LSTM is designed to avoid the problem of long-term dependence. By default, it can remember information for a long time. All RNNs have the form of a repeating module chain, whose structure in a standard RNN is very simple, such as a single tanh function.

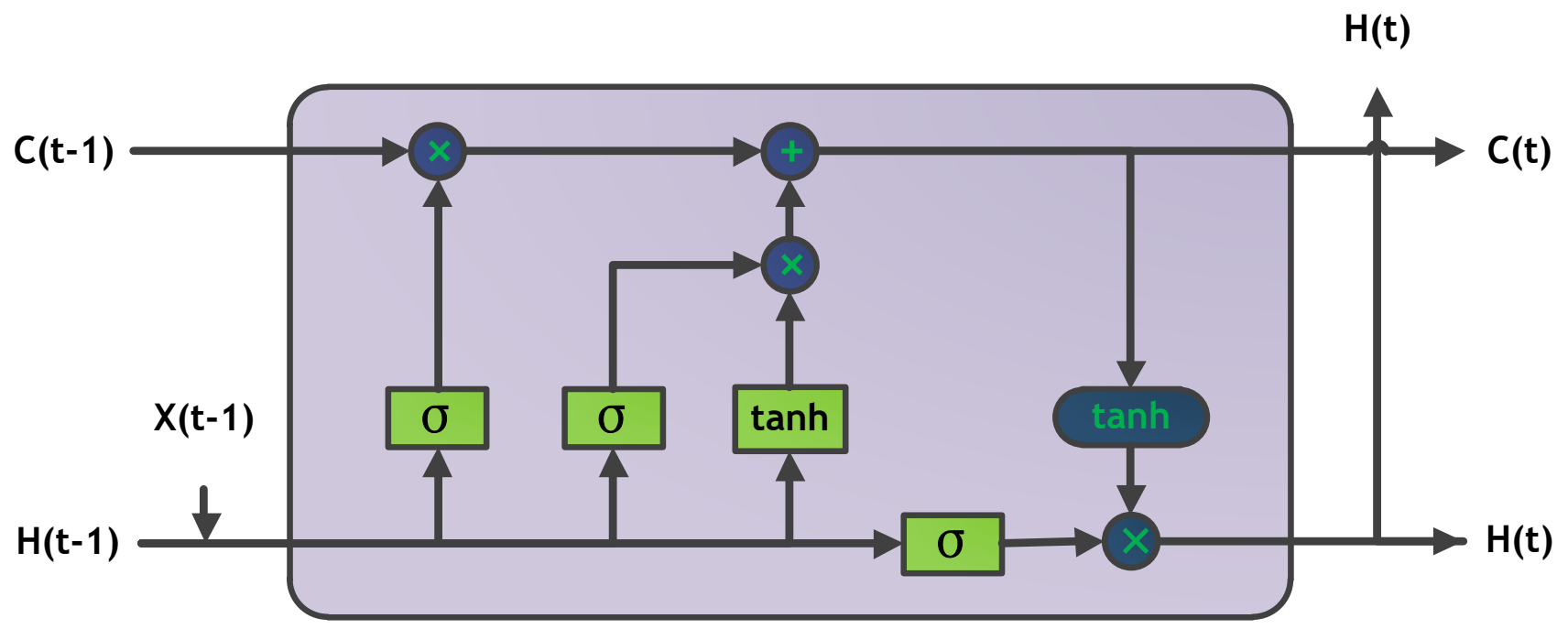

**Fig**. 2 The architecture of LSTM

LSTM also has a similar chain structure as a standard version of RNN, but the repeating module has a different structure as is illustrated in **Fig**. 2. LSTM has three gates: forget gate (input gate layer), input gate layer (input gate layer), and output gate (output layer) to protect and control the state of neurons. The output of the sigmoid activation function is between 0 and 1, indicating how much of each component is allowed to pass. A value of 0 indicates that no content is allowed to pass, and a value of 1 indicates that all content is passed. In addition, there are varied versions of this model.

A popular variant of LSTM proposed by Gers et al. [22] in 2000 is adding peephole connections. It is necessary to use the gate structure to view the neuron status. In this version, peepholes are added to all doors. Another version is to use the coupled forget gate and input gate to make a decision together instead of separately determining which old information to forget and what new information should be added to it, which means that only when we input some information will we forget some older information, and only by forgetting these older information will we input some new information into the neuro state..

LSTM can handle with sequences but lack of the ability to processing long-length data inputs in a fast speed. Compared with it, Dauphin et al. [24] proposed a novel language model based on convolutional neural network, called gated convolutional neural network (Gated CNN). Compared with traditional convolutional neural network, it only adds gates to the convolutional layer. The purpose of the control mechanism is to be able to automatically extract more abstract features. The model consists of multiple layers, similar to the grammatical form of a classic CNN, this model is capable of hierarchical analysis of input, adding a syntactic tree structure and increasing granularity. Compared with RNN that can merely process input sequences one by one, CNN can process input sequences in parallel, which greatly speeds up training. Moreover, the hierarchical structure also simplifies learning, reduces the number of non-linear calculations, thereby alleviating the problem of gradient disappearance and accelerating the convergence rate of the model.

As can be seen from the above, CNN and Gated CNN models cannot provide interpretability as well as semantic similarity, which turns to be the weakness of the malware detectors using these models. In our work, we combine the strengths of CNN and LSTM in order to obtain the relatively better prediction results.

### 2.3 DL-based detection methods overview

Yuan et al. [11] used a combined dynamic and static analysis method to extract a total of more than 200 API call sequence features, and then used feature fusion as input to a deep belief network to build a classification model. However, although their work involved with hybrid analysis, the data input preprocessing stage is not as effective as our method called Glove.

Moreover, DL-based detection can directly use the original binary files without extraction stage, but it is lack of interpretability. For instance, Saxe et al. [12] proposed a method for building a detection model of 400,000 malware original binary files based on deep neural networks, which is free of malware semantic analysis but highly lack of dependability and trust. Hence, our work does not give up the malware information extraction and selection though it can cost some resources.

Although these features selection methods seem inevitable so far, to choose the effective semantic features also brings challenges. For instance, Fan et al. [13] extracted the malware instruction opcode, proposed a sequence mining algorithm to discover the malware opcode sequence pattern, and built a classifier based on an artificial neural network to do an effective detection. Next, Pascanu et al. [14] extracted the semantic system call sequences of malware and detected and classified them based on recurrent neural network (RNN). The experimental results showed that GRU (Gated Recurrent Units) [15] and LSTM [16] models all show excellent classification performance. As mentioned above, opcode and system calls are as important as API calls in mitigating cyber-attacks. In addition, Zhang et al. [18] proposed a method of extracting semantic features using abstract syntax trees and a malware detection method of BP neural network, which has a high recall rate and accuracy rate.

The aforementioned work both use DL-based models and feature extraction techniques. Although this combination can promise trust and prediction accuracy to some degree, its data extraction stage still costs too much computation resources and time and cannot be suitable for malware variants. A novel approach to converting binary files into images so as to get rid of the nature of expertise has been a trend recently. Cui et al. [17] expressed the features of the malware samples as grayscale images and used CNN to automatically extract the image features generated from datasets to identify and classify them. Although the image-based malware detection is creative, we were unable to fully evaluate its accuracy and interpretability through experiments due to that the current work has only explored the local interpretation after the prediction results, which indicates us that image-based malware detection can get high prediction accuracy but lack of global interpretation as an insight given into the models. In another word, image-based models are less interpretable than semantic sequences-based models. Hence, we choose semantic sequences to boost our detection models.

In general, from the literature and related work, we can see that the DL-based malware detection methods, using RNN or CNN alone have some shortcomings. For instance, the length of the feature sequence is not fixed, so when the recurrent neural network LSTM is used alone as the detection model of the malware, the LSTM model cannot extract sequence feature information of too long length. When detecting a classification model, after CNN training, the features do not have contextual relevance and similarity to give model interpretation, so the detection effects may be affected. To address this issue, we provide a novel approach to combining LSTM and CNN, fusing features of APIs and opcode based on Glove vectorization, and evaluating and comparing the results through several experiments.

## 3 The enhanced malware detection approach

This section introduces a methodology for using Glove vectorization to preprocess features generated from malware samples, using MCC_RCNN model to classify and predict malware families and using feature fusion techniques to improve the prediction accuracy. Note that we only use static analysis features due to that the dynamic analysis can cost lots of resources and sometimes cannot trigger the entire malicious behaviors under the virtual environment.

## 3.1 Malware feature extraction and vectorization

We implement a vectorized representation model of malware features based on Glove as is illustrated in **Fig**. 3.

The traditional instruction layer and semantic layer feature representation methods lack the description and similarity of malware context behavior information, so this section proposes a Glove-based feature vectorized representation model. In addition, we use a specific version of LSTM for prediction and evaluation.

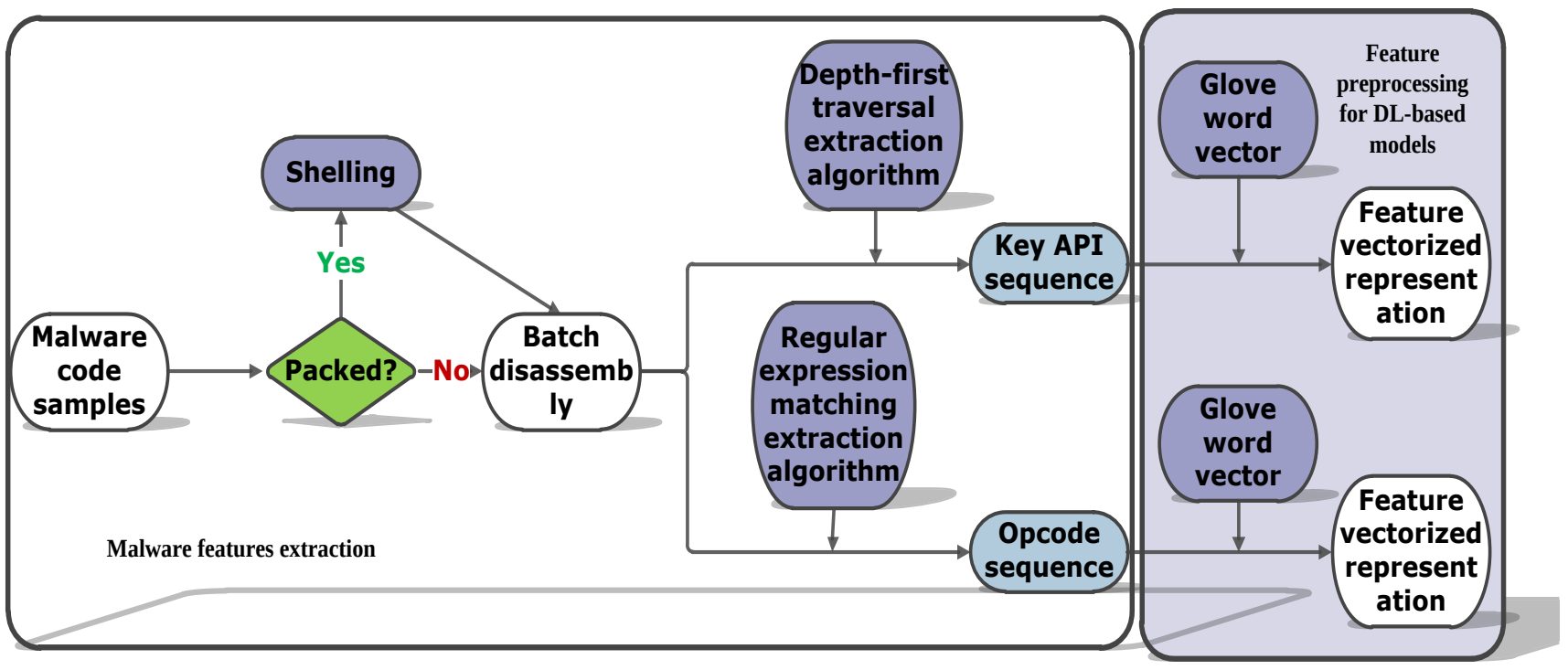

**Fig**. 3 Malware vectorized representation process workflow

### 3.1.1 Feature extraction

Feature extraction mainly include instruction layer and semantic layer extraction.

There are two key steps in the instruction layer feature extraction, which are batch disassembly and feature extraction. We first use IDA Pro to disassemble malware samples and generate files with the suffix ASM in batches. Then we have to extract features from the ASM files. Usually there are four key sections in disassembled batches: text section, idata section, rdata section, data section, where the text section is the program code section, and the instruction opcode all appear in this text section. We extract each line of each ASM file with .text instructions in the same order as the original one. Each malware forms a text file, and each line represents all executable instructions of an ASM file. The core algorithm is shown in **Table** 1:

**Table** 1 Operation code sequence extraction core algorithm

| Algorithm: opcode sequence extraction algorithm based on regular expression matching |
| --- |

Input: Batch disassembled .ASM file
Output: opcode sequences opcode-sequence
Steps：
1.Define a list named opcode-sequences, used to store the opcode sequence
2.p=re. compile(r'\s([a-fA-F0-9]{2}\s)+\s*[a-z]+)')
3.open one .ASM file
4.for line in file: //read the file through the lines
5.if line. Start-with (".text" || ".CODE")//whether 'text' or '.CODE' have begun
6.m = re. find-all (p, line)
    7.Determine whether the content of the regular expression match is found
    8.If found, the opcode = n [0] [1]
    9.If the opcode ! = "Align"
    10.Add a matching opcode to the opcode-sequences list
    11.return opcode sequence opcode-sequence

The semantic feature extraction layer is to extract the key API sequences from the disassembly result using statistical model language algorithm. API refers to a pre-defined function, so that developers do not need to understand its source code. Hence, we can use a group of threads and understand how it works internally by the name of the API. Given the huge cost generated from dynamic analysis, we only extract the static APIs. The strength of this is that it represents the Windows APIs that will be used by the entire instruction path of the malware program. This process has two key steps. In the beginning, we build the key API library of malware, using the disassembly results to build the key API relationship flow graph, and then we use the depth-first traversal algorithm to traverse the key API relationship flow graph to extract the key API call sequence. The key algorithm is shown in **Table** 2.

**Table** 2 The key API sequence extraction algorithm list

Algorithm: Key API sequence extraction algorithm based on depth-first traversal

Input: Disassembled .ASM file
Output: API sequence API sequences
Steps:
1.Calculate the address address−entry at the entrance, code start address add−begin and code end address add−end
2.while current instruction address <add−end do
3. if current API invocation command then
  4. Store system invocation address and type information
    5. else if the current instruction is a jump instruction (unconditional jump instruction, conditional jump instruction, process jump instruction)
    6. then Store the start address, end address and other information of the jump instruction
    7. else if the current instruction is a custom function call instruction then
    8. Store function call start address, end address and other information

9. end if
10.end while
11.Establish a store-call relation dictionary dic = {}
12.for each jump instructions and custom instructions do
13. Linear scan
14. Store the jump relationship, call API relationship, call relationship in dic
15.end for
16.Starting from add-begin, use the depth-first traversal algorithm to call jump instructions and custom functions to extract key API sequences
17.return the key API sequence

3.1.2 Glove-based features vectorized representation

Glove method takes the opcode sequence and the API sequence as inputs, and represents the similarity between them by the spatial distance between the word vectors. In addition, Glove combines the strengths of statistical information of the global vocabulary cooccurrence and the local window context methods such as n-gram window sliding extraction and it does not need to calculate those cooccurrences to 0 so as to greatly reduce the amount of computing and storage.

In the preprocessing stage of our approach, we do not use Word2Vec but Glove as our models for several following reasons: according to the related work investigated before, it is proved that the performance of Glove is far superior to word2vec. Moreover, Glove word vector model is easier to parallelize, which means that for larger training data, Glove is faster. Given the case that the basic dataset used in this experiment is a tremendous number of malware binary files, we choose Glove model to vectorize the malware features.

After obtaining the opcode sequences, we proceed to vectorize them. The corpus has a unique number corresponding to each word, called the word number. First, the opcode sequence is converted into a word number sequence. Then, the words represented by the word number sequence are represented as Glove word vector form. Through these two steps, the opcode sequence extracted from the malware dataset can be represented as a feature map.

The opcode sequence vectorization model represents each opcode sequence as a k-dimensional real vector. Therefore, the n opcodes are converted into n k-dimensional real vectors, that is, $n \times k$ real matrices. Each malware sample is transformed into an $n \times k$ real matrix. Among them, n represents the number of opcodes included in the opcode sequence that symbolizes the entire sample. k indicates that each malware opcode is represented as a k-dimensional real vector. The vectorization process of the malware opcode sequence is shown in the **Fig**. 4(a).

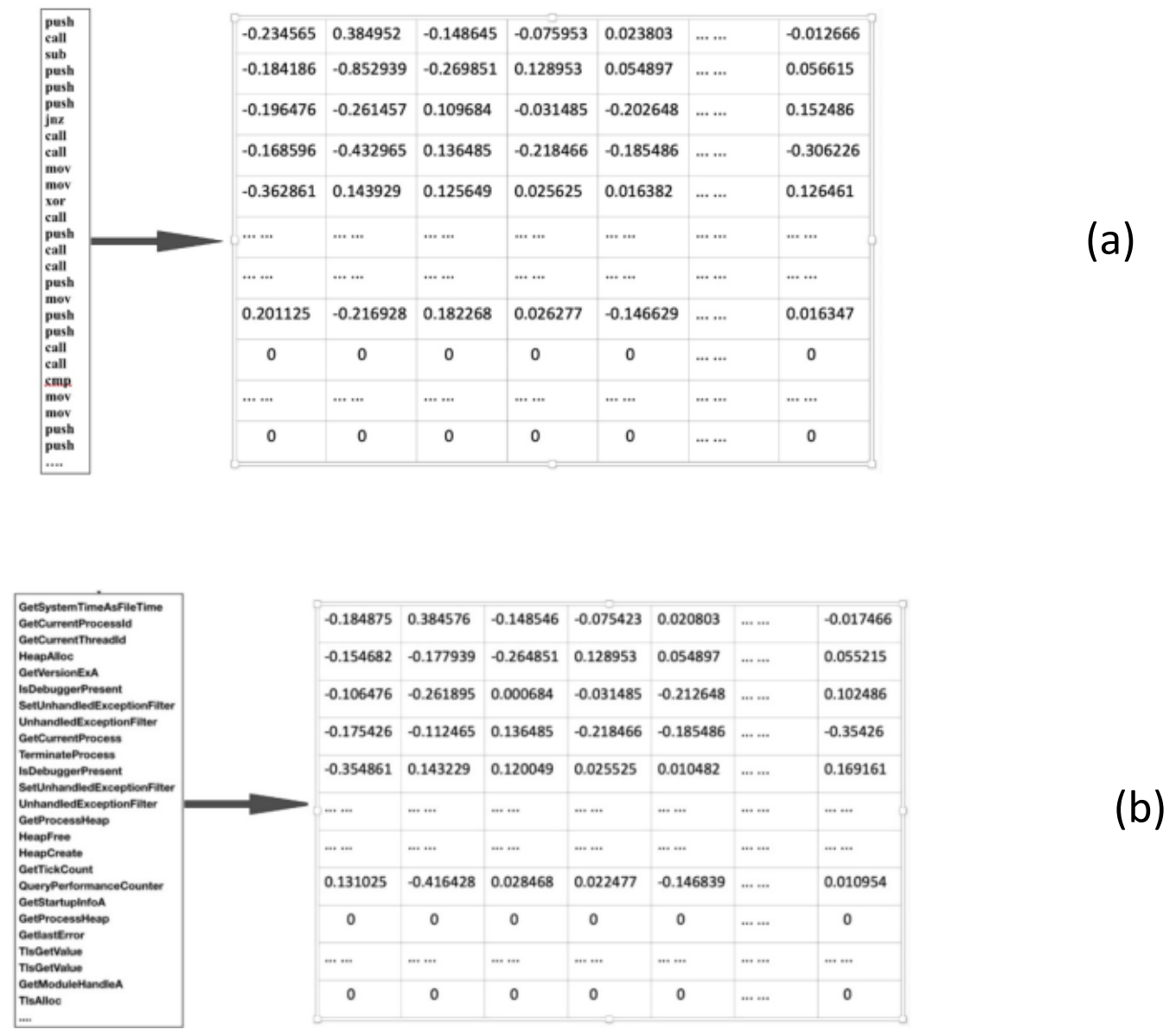

**Fig**. 4 Feature preprocessing: opcode and API

After obtaining the text information of the key API sequences, we perform the following process flow to represent them in vectors. The process is the same with the case in the previous section using opcode. First, the key API sequence is converted into a word number sequence, and then the words represented by the word number sequence are expressed as Glove word vector form, and the key API sequence can be expressed as a feature map. Malware code API sequence vectorized processes are shown as follows with **Fig**. 4(b).

In the model training stage, we propose a single-layer LSTM unit applied to construct a malware detection model, and a structure composition model with two opposite directions is selected, because the LSTM model can extract deeper features of malware due to that it has the preservation mechanism, forgetting mechanism and long-term memory information. After the LSTM model is connected to the softmax classifier, the deeper features can be automatically extracted. The feature vectorization matrix based on Glove and the vector feature based on n-gram are used as the input of LSTM, and the model is trained, adapted to the input data by adjusting the parameters, and an experimental comparison is made to verify our method outcomes.

### 3.2 MCC_RCNN malware detection model

When using a recurrent neural network (such as LSTM) alone as a detection model for malware, the LSTM model cannot extract sequence feature information whose length is too long. When the CNN model is used alone as a detection model for malware, after training, the features do not have contextual relevance and similarity. Hence, we propose a combination model, named MCC_RCNN (Malware Detection_ Recurrent Convolutional Neural Network). First, we input the opcode feature sequence of the malware into the LSTM, obtain the long-sequence operation behavior information through the LSTM, and then input it into the Gated CNN to extract local features of different dimensions. The process of sequence feature information can also eliminate the problem of the opcode association where the CNN model is applied alone or the feature processed by the CNN model has no context.

We separately matrix semantic feature vectors and instruction feature vectors, and feature matrices with sequence features and semantic features are used as LSTM input. Then there appears an individual feature cyclic network design format, which can be used to extract deeper features from the sequences, and put the same length input generated from the hidden layer as the Gated CNN output. Next, we use the output of LSTM as the input of Gated CNN for further feature extraction.

In general, we combine LSTM and Gated CNN to detect and classify malware samples. The MCC_RCNN model combined with malware detection is shown in **Fig**. 5.

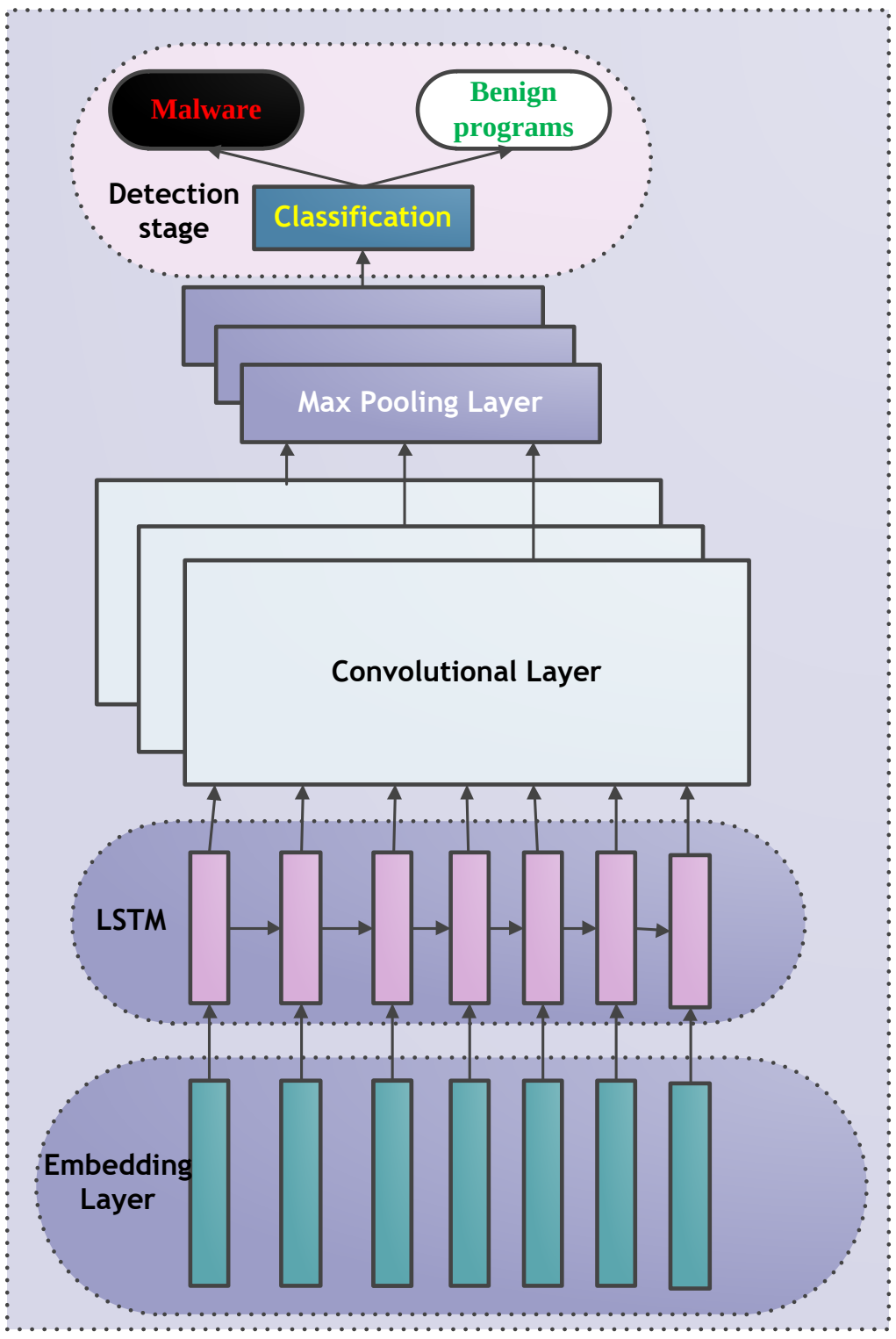

**Fig**. 5 MCC_RCNN model for malware detection

### 3.3 Feature fusion-based malware detection

From section 3.2 to section 3.3, we use single-level features to investigate malware detection models. However, the static features at a single layer can only represent malware to a limited degree. In order to improve the feature description ability and make the malware being described more comprehensive, we fuse the instruction layer and semantic layer features together. The core algorithm of feature fusion used in our work is shown in **Table** 3.

**Table** 3 Feature fusion algorithm list

| Algorithm: Feature Fusion Core Algorithm |
| --- |
| Input: feature f_1 and f_2<br>Output: feature after fusion called f<br>Steps：<br>1. read f1 (the first feature)<br>2.read f2 (the second feature)<br>3.read labels (labelled files)<br>4.use pandas. merge () and malware code id to fusion both the two features<br>5.use pandas. merge () and malware code id to fusion both the features and labels<br>6. return Labeled fusion features |

Similar to the matrixing of the single feature sequence, this part fuses Glove-based malware instruction layer and semantic layer feature vectors into a vector matrix. After batch disassembly and opcode feature extraction algorithms, the opcode sequence and API sequence that can represent the behavioral features are extracted from the binary files, and each opcode and API is expressed as a dimensional real vector. Hence, these opcode sequences of total for n are converted into n-dimensional real vectors, that is, $n \times m$ real matrix; a malware API sequence is also converted into $n \times m$ real matrix. Using the method of dimensional stitching, the two matrices are combined into an $n \times 2m$ fusion matrix.

# 4 Experiments

In the experiments, we provide several comparative experiments for evaluating our approach.

## 4.1 Datasets and experimental setting

The experimental dataset in this paper comes from the dataset of a malware detection competition (Malware Classification Challenge) launched by Microsoft on Kaggle in 2015, the world's largest data modeling and data analysis competition platform. The total size is 136GB, including 9 families of malware. These families are shown in the **Fig**. 6.

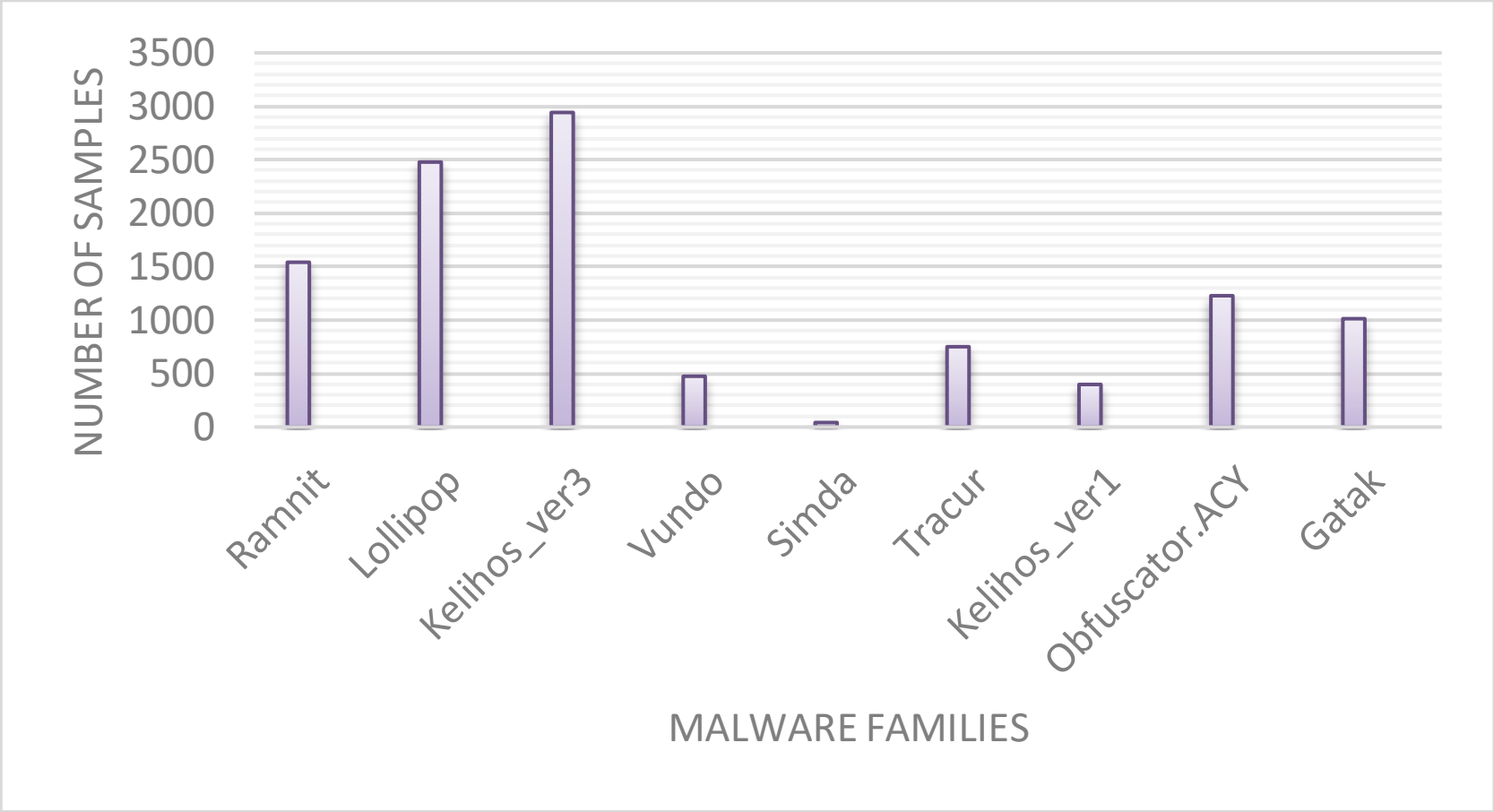

**Fig**. 6 Malware datasets used in this paper

In this paper, a 10-fold cross-validation method is used to prevent overfitting and improve the classification accuracy of all malware data sets. The ratio of training set to test set is 9: 1.

The experimental tools include IDA Pro7.0, the most powerful disassembly tool for static analysis, and shell inspection tool PEID, the shelling tool universal shelling tool. The GPU has 4 cores and is GTX 1080Ti. Memory is 32GB, and disk storage is 220GB SSD with 3TB hard disk. The program language we use to accomplish the algorithms is Python 3.6 whereas we use Tensorflow 2.0 for deep learning model training.

### 4.2 Evaluation indicators

Aiming at verifying the malware classification model, accuracy, precision, and F1-Score are used as evaluation indicator, which are calculated based on TP, TN, FP, FN. Since 9 types of malware are used in our work, it is a multi-classification problem. In the binary classification model, the confusion matrix is constructed according to True / False and Positive / Negative. The accuracy rate is used to evaluate the correct rate detection model, and is the most intuitive evaluation indicator. Accuracy is defined as the ratio of the number of samples correctly classified by the classification model to the total number of samples for a given test dataset.

The accuracy rate refers to the proportion of true malware (TP) out of all malware samples (TP + FP). The calculation formulas are the formulas (1) as follows, where l represents the number of categories:

$$\text{Accuracy} = \frac{1}{l}\sum_{i=1}^{l}\frac{TP_i + TN_i}{N} \quad (1)$$

### 4.3 Glove-based vectorization comparative experiment

In order to verify the effectiveness of the feature representation method based on Glove word vectors, we compare the opcode sequence matrix generated from Glove word vectorization and the opcode vectorized features generated from n-gram window sliding approach after being trained by LSTM models for detection, which is labelled as comparative experiment A.

When the instruction layer uses n-gram algorithm window sliding to obtain features, after the LSTM model is successfully established, in order to adapt the model to malware data inputs, it is necessary to adjust the parameters of LSTM and optimize the network structure of LSTM. Hence, we need to tune the model first. The parameters such as the number of test iterations epoch-size, the learning rate l-r, and the number of hidden neurons n-hidden affect the classification results of the LSTM malware detection model. The value of n in the n-gram algorithm is also worth studying. On these issues, n is fixed at 4, and after several experimental times, the model has a good classification capacity when the learning rate l-r is 0.002, the number of iterations epoch-size is 15, and the number of hidden neurons n-hidden is 20. Set the value of n in the n-gram algorithm to 1 to 4, and test the classification accuracy. The experimental results are shown in **Fig**. 7(a1). It can be seen that when n is 2, the classification accuracy is the highest, 77.38%.

When the 300-dimensional opcode word vector is used as the input of the LSTM, the performance capabilities of the opcode are different under the combination of different parameters of the LSTM. After several experiment times, a set of optimal parameter combinations is obtained. When the learning rate is 0.001, the number of iterations epoch-size is 20, and the LSTM dimension is 128 dimensions, the model has the highest accuracy and can adapt well to the sample features. From the comparative experiment results in **Fig**. 7(a2), it can be seen that the accuracy of the Glove-based feature representation model is higher than that of the traditional n-gram feature extraction method. In the family classification problem, the generalization ability of the model is determined by the above indicators, but mainly by accuracy to determine the classification effect of the model. The accuracy of the Glove method is 83.95%, which is 6.57 percentage points higher than that of the n-gram algorithm when n is 2 with better performance, indicating that the description capability of the Glove-based feature vectorization method is better.

From the Experiment A results comparison, we can see that the vectorized representation method of the opcode sequence feature based on the Glove word vector is better than the method based on the traditional n-gram and the method based on Glove can use the spatial distance between the opcode sequence and the API sequence to express the correlation and similarity between the contextual opcodes so as to describe the malware from the behavior level more accurately.

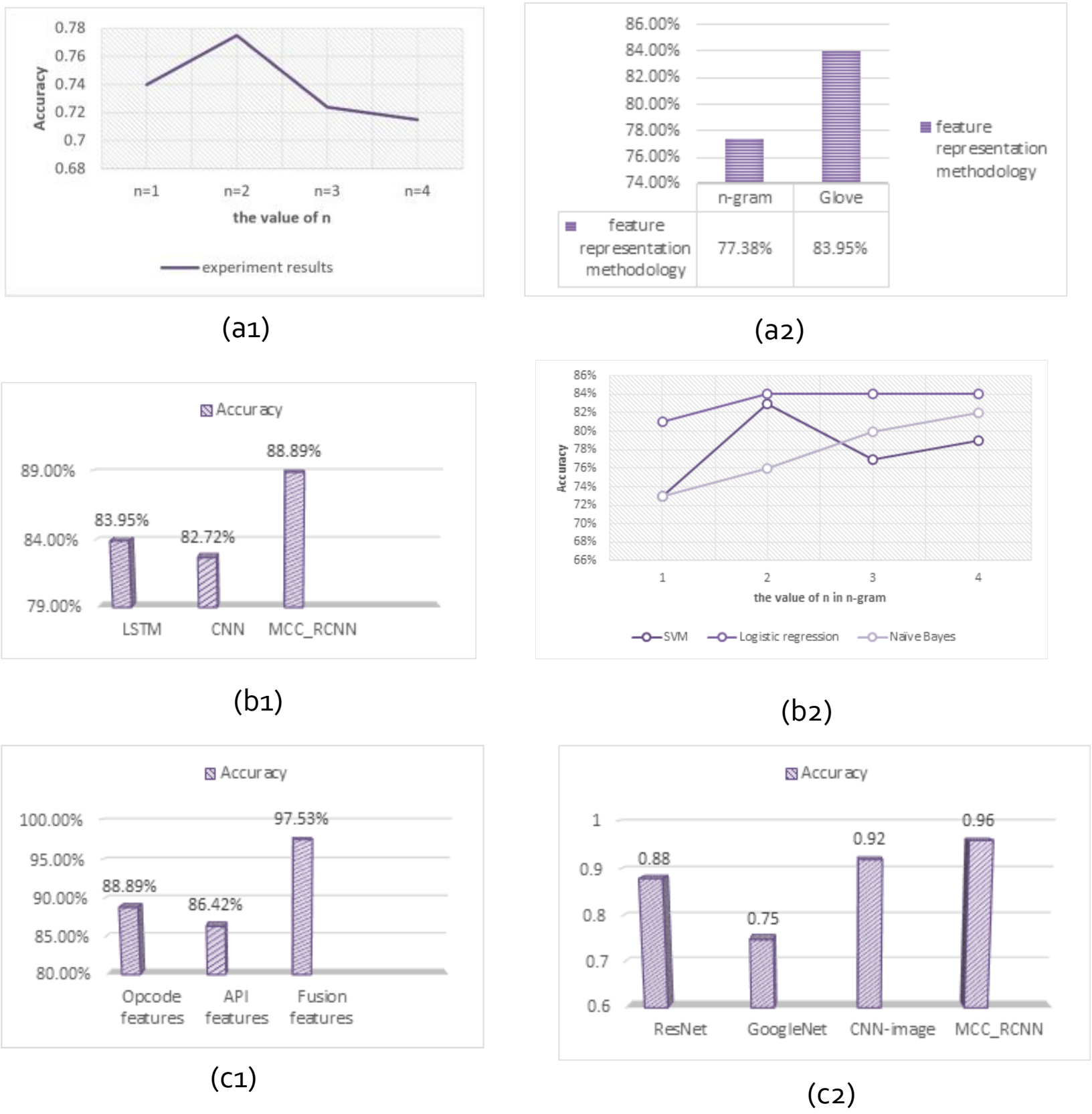

**Fig**. 7 Experimental results and analysis

## 4.4 MCC_RCNN-based malware detection comparative experiment

We set up the comparative experiment B1 to verify the effect of the MCC_RCNN model proposed. In this comparative experiment, we input opcode sequence matrix into three models: LSTM model, Gated CNN model, and MCC_RCNN model. After training and detection, we compare the results.

By comparing experiment B1 as is illustrated in **Fig**. 7(b1), we can know that the accuracy of the MCC_RCNN detection model proposed in this section is higher than other models. It is verified that the MCC_RCNN can synthesize the advantages of CNN and RNN, and the effect is better than the classification accuracy of traditional ML classifier SVM model, recurrent neural network LSTM model and CNN model.

On the other hand, to test whether the classification effect based on Glove's feature vectorization method and MCC_RCNN detection model is better than the classification effect of traditional ML, a comparative experiment B2 is set. We first input Glove-based feature vector matrix into MCC_RCNN model, then input n-gram extraction feature vectors into Logistic regression, Naïve Bayes, and KNN respectively. The n of the n-gram algorithm is fixed at 4 and the frequency is fixed at 700.

From **Fig**. 7(b2), we can know that when n has a value from 1 to 4, three ML-based classification algorithms have different performance and when n equals 2, there comes the highest accuracy score in SVM and Logistic regression. In addition, when n equals 4, the accuracy score obtained by Naïve Bayes classifier comes the highest. However, the accuracy score coming from Glove-based representation using MCC_RCNN is 88.89%. This outcome is better than any one of three cases shown in the above figure (note that the highest shown above is 84%), which indicates to us that our work comprising of Glove and MCC_RCNN is better at predicting malware samples.

### 4.5 Feature fusion-based malware detection comparative experiments

In order to test whether the classification effect of feature fusion is better than that of a single local feature, the instruction layer feature vector matrix and the semantic layer feature vector matrix are fused. We set up comparative experiment C, where we first input API sequence matrix and opcode sequence matrix into MCC_RCNN model and then input fusion feature matrix into MCC_RCNN model. After training, we can compare and analyze the results under these three situations.

As shown in the experimental results in **Fig**. 7(c1), the classification accuracy based on API features is 86.42%, the classification accuracy based on opcode features is 88.89%, and the classification accuracy based on fusion features is 97.53%, which is significantly higher than a single semantic layer. The accuracy rate of API features and instruction layer opcode features shows that the feature description method that combines the instruction layer and semantic layer features is better than the single feature description method. It verifies that the fusion feature describes the behavioral information of malware from different dimensions and improves the ability to describe the behaviors more comprehensively, while single-level features can only describe the behavior information from one dimension.

Next, we compare the effect of the malware detection model based on Glove and MCC_RCNN proposed in this paper with the literature [27] [28] to verify the effectiveness of the detection model proposed in this paper. Note that these works use the same malware datasets as ours. In [27], researchers extracted the opcode sequences and transformed them into images, and analyzed the classification results of the images on the ResNet and GoogleNet models respectively. In [28], researchers mapped the byte sequences into color images and classifies them using the classic CNN model.

According to the analysis on the experimental results in **Fig**. 7(c2), the classification accuracy of the fusion feature is 97.53%, which is higher than the accuracy of the literature [27] using the ResNet model and GoogleNet model, and also higher than the accuracy of the paper [28] classic CNN model. The detection model based on feature fusion in our work is better than the work in literature. The classification model using RNN or CNN alone is better. The effectiveness of the fusion feature detection model proposed in this chapter is verified.

## 5 Conclusions and future work

The detection of malware has great research significance and application value. DL-based methods are the current trend in the field of malware detection. Based on the research of current malware detection methods, a vectorized representation method of malware instruction layer and semantic layer based on Glove is proposed in this paper. Moreover, we fuse the features from the instruction layer and the semantic layer, and provide a more accurate feature description method based on the feature fusion. In addition, we also implement a malware detection model MCC_RCNN based on CNN and RNN, which uses LSTM to save the complex logical relationship between feature sequences, and uses Gated CNN to extract features of different dimensions. From Glove vectorization and MCC_RCNN, to the techniques of feature fusion, we implement several comparative experiments to evaluate the effectiveness of these work. Hence, not only can this approach solve the problem that the LSTM model cannot extract sequence feature information of too long length, but also it can eliminate the problem that the application of the CNN model alone or the features processed by the CNN model no longer have contextual opcode association. Compared with traditional machine learning classification models and traditional CNN and RNN models, the designed mechanism in this paper is verified to improve the accuracy of detection and classification.

Although we propose a feature selection method that has not been tried and has a good effect, many works and research contents are worth further discussion and research:

(1) We choose to extract features from two levels of the malware's byte layer and instruction layer and then conduct the related experiments based on static analysis. Because static analysis relies heavily on the results of using disassembly tools, and the major issue of disassembly is that it will be affected by code obfuscation, encryption, and compression, thus only the features generated from malware are extracted from these static levels. Hence, this analysis is worth further discussion and being considered with implementing dynamic analysis or hybrid analysis.

(2) we can add attention mechanism to the vectorized expression of malware features. This article is based on Glove's vectorized representation, but it lacks judgment on the importance of malware behavior features and research on independent behaviors. Moreover, awareness of automatically recognizing the API function one by one in a sample hasn't been fully explored yet. Later, we can give improving the model based on the attention mechanism to make the characterization of malware more comprehensive into consideration.

## Availability of data and material

Not applicable.

## Funding

Not applicable.

## Acknowledgements

Not applicable.